\documentclass[fleqn,twoside]{article}
\usepackage[headings]{espcrc2}

\readRCS
$Id: espcrc2.tex,v 1.2 2004/02/24 11:22:11 spepping Exp $
\usepackage{graphicx}
\usepackage[figuresright]{rotating}

\newcommand{\AmS}{{\protect\the\textfont2
  A\kern-.1667em\lower.5ex\hbox{M}\kern-.125emS}}
\newcommand{\NeqFour}{{\cal N}=4}

\def\fig#1{fig.~{\ref{#1}}}

\def\eqn#1{eq.~(\ref{#1})}

\def\Ksl{\s{K}}
\def\sandp#1.#2.#3{%
\left\langle\smash{#1}{\vphantom1}^{-}\right|{#2}%
\left|\smash{#3}{\vphantom1}^{+}\right\rangle}
\def\sandpp#1.#2.#3{%
\left\langle\smash{#1}{\vphantom1}^{+}\right|{#2}%
\left|\smash{#3}{\vphantom1}^{+}\right\rangle}
\def\sandmm#1.#2.#3{%
\left\langle\smash{#1}{\vphantom1}^{-}\right|{#2}%
\left|\smash{#3}{\vphantom1}^{-}\right\rangle}
\def\spab#1.#2.#3{\sandmm#1.#2.#3}
\def\spba#1.#2.#3{\sandpp#1.#2.#3}
\def\spaa#1.#2.#3.#4{\sandmp#1.{#2#3}.#4}
\def\spbb#1.#2.#3.#4{\sandpm#1.{#2#3}.#4}
\def\spa#1.#2{\langle#1\,#2\rangle}
\def\spb#1.#2{[#1\,#2]}
\def\spash#1.#2{\vphantom{\hat K}\spa{\smash{#1}}.{\smash{#2}}}
\def\spbsh#1.#2{\vphantom{\hat K}\spb{\smash{#1}}.{\smash{#2}}}
\def\lor#1.#2{\left(#1\,#2\right)}
\def\sand#1.#2.#3{%
\left\langle\smash{#1}{\vphantom1}^{-}\right|{#2}%
\left|\smash{#3}{\vphantom1}^{-}\right\rangle}
\def\sandpp#1.#2.#3{%
\left\langle\smash{#1}{\vphantom1}^{+}\right|{#2}%
\left|\smash{#3}{\vphantom1}^{+}\right\rangle}
\def\sandpm#1.#2.#3{%
\left\langle\smash{#1}{\vphantom1}^{+}\right|{#2}%
\left|\smash{#3}{\vphantom1}^{-}\right\rangle}
\def\sandmp#1.#2.#3{%
\left\langle\smash{#1}{\vphantom1}^{-}\right|{#2}%
\left|\smash{#3}{\vphantom1}^{+}\right\rangle}
\def\eps{\epsilon}
\def\Res{\mathop{\rm Res}}

\def\NeqFour{{\cal N}=4}

\newcommand{\nn}{\nonumber}
\def\cg{c_\Gamma}
\def\Remaining{{\widehat {R}}}
\def\Cuth{{\widehat {C}}}

\newbox\SlashedBox
\def\slashed#1{\setbox\SlashedBox=\hbox{#1}
\hbox to 0pt{\hbox to 1\wd\SlashedBox{\hfil/\hfil}\hss}#1}
\def\hboxtosizeof#1#2{\setbox\SlashedBox=\hbox{#1}
\hbox to 1\wd\SlashedBox{#2}}

\newbox\charbox
\newbox\slabox
\def\s#1{{      
        \setbox\charbox=\hbox{$#1$}
        \setbox\slabox=\hbox{$/$}
        \dimen\charbox=\ht\slabox
        \advance\dimen\charbox by -\dp\slabox
        \advance\dimen\charbox by -\ht\charbox
        \advance\dimen\charbox by \dp\charbox
        \divide\dimen\charbox by 2
        \raise-\dimen\charbox\hbox to \wd\charbox{\hss/\hss}
        \llap{$#1$}
}}

\title{On-Shell Unitarity Bootstrap for QCD Amplitudes\thanks{Presented at Loops \& Legs 2006, April 23--28, 2006, Eisenach, Germany}}

\author{Carola F. Berger\address[SLAC]{
        Stanford Linear Accelerator Center \\
        Stanford University\\
        Stanford, CA 94309, USA}%
        \thanks{Research supported in part by the US Department of
                Energy under contract DE--AC02--76SF00515},
        Zvi Bern\address[UCLA]{Department of Physics and Astronomy, UCLA\\
               Los Angeles, CA 90095--1547, USA}%
\thanks{Presenter. Research supported by the US Department of
        Energy under contract DE--FG03--91ER40662},
        Lance J. Dixon\addressmark[SLAC],
        Darren Forde\address[Saclay]{Service de Physique Th\'eorique,\\ 
                         CEA--Saclay, F--91191 Gif-sur-Yvette cedex,\\ France},
   David A. Kosower\addressmark[Saclay]\thanks{Presenter} }
       
\runtitle{On-Shell Unitarity Bootstrap for QCD Amplitudes}
\runauthor{Berger, Bern, Dixon, Forde, \& Kosower}

\begin{document}

\begin{abstract}
\vspace{1pc}
\end{abstract}

\maketitle

\thispagestyle{myheadings}
\markboth{}{\rm UCLA/06/TEP/22 \hfil SLAC--PUB--12064 \hfil Saclay/SPhT--T06/096}
\section{Introduction}

\begin{figure}[tbh]
\vspace{9pt}
\vskip -.5 cm 
\begin{center}
\includegraphics[width=8pc]{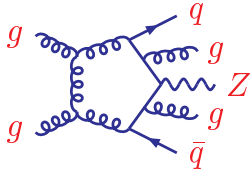}
\end{center}
\vskip -.7 cm 
\caption{An important background to searches for missing energy
signals of supersymmetry and other models of physics beyond the
Standard Model is $pp \rightarrow Z +$ jets.  This calculation
involves seven-point one-loop amplitudes.}
\label{Z4jetsFigure}
\end{figure}

Seeking and measuring new physics at the imminent Large Hadron
Collider (LHC) will require extensive calculations of
high-multiplicity backgrounds in perturbative QCD to 
next-to-leading order (NLO).
The Les Houches 2005 workshop defined a target
list, reproduced in table~\ref{LesHouchesTable}, for theorists
to attack.  In addition to the processes in the table,
one would also like to compute processes such as $W,Z+4$ jets,
which are important backgrounds to searches for
supersymmetry and other models of new electroweak physics. Such
computations require one-loop amplitudes with seven external
particles, including the vector boson, as depicted in
\fig{Z4jetsFigure}.  These are challenging calculations and
Feynman-diagrammatic computations have only recently reached six-point
amplitudes~\cite{FeynmanSix}.  (Some of this progress has been
described in this conference~\cite{OtherLoopLegs}.)

\begin{table}[htb]
\begin{center}
\caption{The NLO target list.  (From ref.~\cite{LesHouches}.)
\label{LesHouchesTable}}
  \vspace*{1mm}
\begin{tabular}{|l|l|}
\hline
&\\
process&relevant for\\
($V\in\{Z,W,\gamma\}$)&\\
\hline
&\\
1. $pp\to V\,V$\,jet &  $t\bar{t}H$, new physics\\
2. $pp\to t\bar{t}\,b\bar{b}$ &  $t\bar{t}H$\\
3. $pp\to t\bar{t}+2$\,jets  &  $t\bar{t}H$\\
4. $pp\to V\,V\,b\bar{b}$ &  VBF$\to H\to VV$,
$t\bar{t}H$, \\
& \hskip .2 cm new physics\\
5. $pp\to V\,V+2$\,jets &  VBF$\to H\to VV$\\
6. $pp\to V+3$\,jets &  various new physics  \\
 & \hskip .2 cm signatures\\
7. $pp\to V\,V\,V$ &  SUSY trilepton\\
&\\
\hline
\end{tabular}
\end{center}
\end{table}

The last two decades have produced an increasing collection of
explicit expressions for amplitudes in gauge theory.  Many of these
results are dramatically simpler in their analytic forms than
would have been expected based on counting Feynman
diagrams and terms therein.  This is especially true in the maximally
($\NeqFour$) supersymmetric theory, but is also true of amplitudes in
QCD, directly relevant to collider experiments.

Many of these results were not obtained using Feynman-diagram
techniques, and some (the one-loop all-multiplicity results, in
particular) are not accessible to calculations done using these
traditional techniques.  
The traditional approach makes the Lagrangian manifestly
symmetric under Lorentz and local gauge symmetries, and hence simple
in form.  The price we pay is the introduction of many 
non-physical degrees of freedom.  To
remove the redundant degrees of freedom in calculations of scattering
amplitudes requires fixing a gauge. This makes computations more
complicated, because individual diagrams do not preserve gauge
invariance, which is recovered only at the end of a long calculation.
We end up calculating a lot of unphysical and redundant information
which is thrown away at the end.  From a practical point of view,
however, what matters more than a simple Lagrangian is simplicity
and efficiency of calculation, where we calculate no more than what 
is really needed for the result.  What we want is a calculational 
formalism that involves {\it only\/} (perturbative) physical states.  
Light-cone gauge is a first step towards this goal, as it
removes the unphysical
helicities propagating inside diagrams.  Like all diagrammatic
formalisms, however, it still involves
off-shell states; in a gauge theory, off-shell formulations are
inherently non-gauge invariant.

What we are seeking is a way of doing field theory in terms of
gauge-invariant, on-shell states.  The possibility of doing this flies
in the face of existing graduate education and a great deal of lore.
Nonetheless, for massless theories we now understand how to do this
explicitly for tree and one-loop amplitudes; and there is every reason to
believe the procedure will work to all orders in perturbation theory.
A theme that runs through the technologies underlying on-shell
methods is the transformation of general properties of amplitudes into
practical tools for computing them.

There are three basic technologies that underlie the on-shell approach
to calculations in gauge theories.  The first is the spinor helicity
method~\cite{SpinorHelicity}, which gives efficient representations
for the physical content of external states.  The second to be
developed was the unitarity method~\cite{BDDK} for obtaining loop
amplitudes with multiple kinematic variables.  This method makes use
of the basic unitarity property of field theory to provide a
systematic procedure for constructing amplitudes with any number of
kinematic invariants.

With the four-dimensional version of the unitarity method one can
compute complete one-loop amplitudes in supersymmetric
theories, and all terms containing branch cuts in QCD or supersymmetric
theories to all loop orders.   Curiously, 
the on-shell method for tree amplitudes, which is the third technology
underlying the on-shell bootstrap approach, was developed well after
the on-shell technique for loop amplitudes.  
The tree-level technique had to await new inspiration from 
the twistor-space picture for amplitudes~\cite{WittenTopologicalString}. 
 The approach at tree level makes use of factorization, another
basic property of field-theoretic amplitudes, to obtain
 on-shell recursion
relations~\cite{BCFRecurrence,BCFW}.  The most recent development,
on which we report here, also gives a method for
computing the rational parts of gauge-theory loop amplitudes, that is
precisely those terms not accessible to four-dimensional unitarity 
(but requiring the computationally more awkward
 full $D$-dimensional unitarity).

A practical method should exhibit only modest growth in its complexity
as the number of external legs increases.  It should lead to 
numerically stable results, allowing for straightforward numerical
integration over all experimentally-accessible phase space.
Conventional methods have neither of these features;
they tend to have numerical
stability issues due to large numbers of high degree spurious
singularities.  (Solutions are being explored, as reported 
at the conference~\cite{FeynmanSix,OtherLoopLegs}.)
The on-shell bootstrap that we describe in this report does;
and it is applicable to a wide class of processes in a manner 
that likely allows for automation.

\section{Hints from Twistor Space}

The recent suggestion, that a twistor-space string theory is dual to
the maximally supersymmetric gauge theory~\cite{WittenTopologicalString}, 
points at
additional structure in scattering
amplitudes.  It meshes nicely with the
techniques we describe and may in the future offer a new formal
framework for these developments.  An important inspiration for
the recent advances comes from the surprising simplicity
amplitudes exhibit in twistor space.

In QCD color-ordering~\cite{Color} and spinor
helicity~\cite{SpinorHelicity} are widely used at tree and loop level
to provide simplified descriptions of amplitudes.  
QCD amplitudes can be expressed entirely
in terms of spinors by representing
gluon polarizations in the spinor helicity basis,
\begin{eqnarray}
\varepsilon^{+}_\mu (k;q) =  {\sandmm{q}.{\gamma_\mu}.k
      \over \sqrt2 \spa{q}.k}\, , \nonumber \\
\varepsilon^{-}_\mu (k;q) =  {\sandpp{q}.{\gamma_\mu}.k
      \over \sqrt{2} \spb{k}.q} \,,
\nn
\end{eqnarray}
where 
\begin{eqnarray}
\epsilon_{a b} \lambda^a_j \lambda^b_l = 
\spa{j}.{l} = \langle j^- | l^+ \rangle = \bar{u}_-(k_j) u_+(k_l)\,, 
\nonumber \\
\epsilon_{\dot a \dot b} \tilde \lambda^{\dot a}_l \tilde \lambda^{\dot b}_j = 
\spb{j}.{l} = \langle j^+ | l^- \rangle = \bar{u}_+(k_j) u_-(k_l)\,, 
\nn
\end{eqnarray}
and $u_\pm(k)$ is a massless Weyl spinor with momentum $k$ and positive
or negative chirality respectively. Lorentz inner products of momenta can 
also be expressed in terms of spinors via
\begin{eqnarray}
s_{ij} = 2 k_j \cdot k_l = \spa{j}.{l} \spb{l}.{j} \,.
\nn
\end{eqnarray}

A twistor-space description arises from 
performing an asymmetric Fourier transform, one with respect to
negative- (but {\it not\/} positive-) helicity spinors,
\begin{eqnarray}
\tilde A(\lambda_i, \mu_i) \!= \!
\! \int\!\! \prod_i {d^2 \tilde \lambda_i \over (2 \pi)^2}
  \exp\Bigl(i \sum_i \mu_i^{\dot a} \tilde\lambda_{i \dot a} \Bigr) 
       A(\lambda_i, \tilde\lambda_i) \,, 
\nonumber
\end{eqnarray}
where the $\lambda_i$ are positive-helicity spinors and $\mu_i$ 
are the conjugate variables to the
negative-helicity ones.  

Witten conjectured that in
twistor space, gauge-theory amplitudes have delta-function support on
curves of degree
\begin{eqnarray}
d = q - 1 + L \,,
\nn
\end{eqnarray}
where $q$ is the number of negative-helicity legs and $L$ the number
of loops. 

\begin{figure}[tbh]
\vspace{9pt}
\vskip -.5 cm 
\begin{center}
\includegraphics[width=15pc]{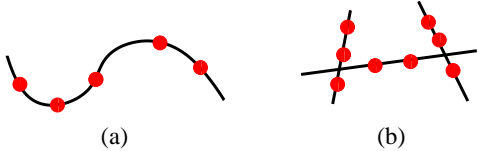}
\end{center}
\vskip -1 cm 
\caption{In twistor space amplitudes have delta-function support
on algebraic curves.  The dots represent the external points.
The curve (a) represents a non-degenerate cubic curve. 
In (b) the curve degenerates to intersecting straight lines.}
\label{TwistorCurveFigure}
\end{figure}

Surprisingly, there are multiple descriptions of the amplitudes in terms
of non-degenerate and degenerate curves~\cite{CSW,RSV}, two of which
are displayed in \fig{TwistorCurveFigure}.  The degenerate description
in terms of intersecting degree-one curves (straight `lines') has been the
most useful for practical calculations.  This description led to the
MHV rules~\cite{CSW} of Cachazo, Svr\v{c}ek and Witten.  They effectively
compute amplitudes in terms of a sum over all multi-particle factorizations.
  The rules provided a concrete demonstration that
scattering amplitudes have a simple underlying structure
not understood previously.

Does this simplicity underlie loop amplitudes as well?  
We have ample evidence that
it does.  In particular, from quadruple cuts one can demonstrate that
in twistor space the coefficients of all box integrals in any massless
gauge theory are have delta-function support on intersecting lines
forming a closed loop~\cite{BCFCoplanar,NeqFourNMHV}. 
This corresponds to the possibility of
computing the coefficients of any box integral
in a 
four-dimensional theory from a product of four tree amplitudes,
by solving four
on-shell constraints~\cite{BCFUnitarity}.  The recent
computation of $n$-point one-loop QCD amplitudes in
refs.~\cite{Bootstrap,FordeKosower,Genhel} also points
at the presence of simple underlying structures.

Witten also made a simple observation
that has proven of great practical utility.  Ordinarily, one works
with real momenta, in which case three-point amplitudes vanish identically.
This results from the vanishing of all momentum invariants $s_{ij}$,
which in turn forces the vanishing of all spinor products $\spa{i}.j$ and
$\spb{i}.j$.  If one were to use complex momenta, however, the vanishing of
momentum invariants only requires one flavor
of spinor product to vanish,
either $\spa{i}.j$ or $\spb{i}.j$.  The other spinor product can be used to 
define a non-vanishing three-point amplitude.  All other amplitudes
can be built out of this basic amplitude.

\section{Unitarity Method}

Our approach is based on unitarity, which has been a fundamental
concept in quantum field theory since since its inception~\cite{Eden}.
In the 60's most attempts to describe the strong interactions relied on
the unitarity and analyticity of the $S$-matrix.  But with the advent
of QCD in the 70's as the description of the strong interactions, Feynman
diagrams became the primary tool for describing 
scattering at large transverse momentum. 

Although an approach based on unitarity offers advantages
because one can avoid using unphysical (off-shell) states, a
number of difficulties prevented its use as a practical tool.
  The primary difficulty was the
inability to obtain amplitudes depending on more than two kinematic
amplitudes via multiple dispersion relations.  Other difficulties
include technical issues in applying unitarity to massless theories,
as well as non-convergence of dispersion relations, which require
subtractions for well-defined results.  A resolution of these
difficulties occurred with the advent of the `unitarity
method'~\cite{BDDK}.

\begin{figure}[htb]
\centerline{\includegraphics[width=15pc]{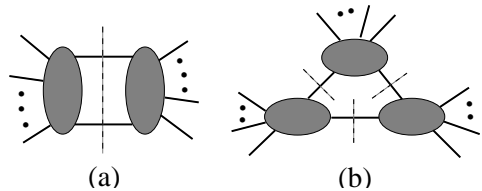}}
\vskip -.7 cm 
\caption{A two-particle cut (a) as well as a 
generalized triple cut (b).}
\label{TripleCut}
\end{figure}

In the unitarity method one systematically constructs amplitudes by
merging the various unitarity cuts as exemplified in \fig{TripleCut}
into Feynman-like integrals which
give the correct cuts in all channels.  In this approach both
dispersive and absorptive parts are obtained simultaneously, bypassing
the need for dispersion integrals.  Furthermore, by making appropriate
use~\cite{vanNeerven}
 of dimensional regularization within the method one can easily
avoid many of the earlier technical complications.

The unitarity approach has proven to be a powerful method for
determining amplitudes, especially in supersymmetric theories where
complete one-loop amplitudes may be obtained using only tree-level
four-dimensional helicity amplitudes as input~\cite{BDDK}.  In
non-supersymmetric theories complete amplitudes may be obtained using
unitarity in $D$ dimensions. The required tree amplitudes in
the latter case are more
complicated, so only a limited number of computations have been performed
with this approach~\cite{BernMorgan}. After performing
a series expansion in $\eps =(4-D)/2$, the difference 
between using four-dimensional or
$D$-dimensional states and momenta in the cuts gives rise
to rational functions of spinor invariants.

Because the scalar bubble, triangle and box integrals form a complete basis of
cut-containing 
functions for dimensionally-regularized
one-loop amplitudes in a four-dimensional theory~\cite{Integrals}, 
all a computation needs to determine
are the coefficients of these
integrals.  Recent improvements to
the four-dimensional unitarity
method~\cite{BCFUnitarity} use generalized
unitarity~\cite{Eden,ZFourPartons,TwoLoopSplit} to allow for a direct
algebraic determination of box integral coefficients.  The
efficiency of extracting the coefficients of bubble and triangle 
integrals has also been improved recently~\cite{BBCFSQCDBFM}, 
and has been applied to six-gluon amplitudes.  
Combined with the previously-computed results for the
cut-containing pieces~\cite{BDDK,BBSTQCD}, these results give a complete
analytic solution for the cut-containing terms in all six-gluon
amplitudes.

\section{On-Shell Recursion Relations}

This leaves us to compute
the rational terms efficiently.  We again use
analytic properties, but instead of branch cuts we use the poles, in
the guise of on-shell recursion relations.  Our focus here will be on
obtaining results with large numbers of external legs. Our interest in
constructing all-multiplicity amplitudes stems partly from the desire
to study the growth in complexity of the amplitudes as the number of
external partons increases.  Because the explosive growth in complexity
with each additional leg has been a stumbling block in previous
methods, it is important to understand this behavior with any new
method.  Furthermore, experience has shown that analytic all-$n$
expressions provide a wealth of information about the general structure
of scattering amplitudes.

The on-shell unitarity 
bootstrap~\cite{Bootstrap,Genhel,LoopMHV} has its origins
in an early approach taken to compute the $Z\rightarrow 4$~parton
one-loop matrix elements~\cite{ZFourPartons} (or equivalently, by
crossing, for $pp \rightarrow W, Z + 2 \,{\rm
jets}$). In this more primitive version of the on-shell bootstrap
approach, the cut-containing (poly)logarithmic terms were obtained using
the unitarity method while purely rational terms were obtained using
on-shell factorization properties, writing down an
ansatz and constraining its form.
It proved difficult to turn the approach into a general and
systematic one.  On-shell recursion for the rational terms
provides such a general and systematic method.

In special cases, when certain criteria are satisfied by the cuts, one
may even use on-shell recursion to obtain the cut-containing terms of
amplitudes~\cite{RecurCoeff}. That is, one may use the kinematic poles
appearing in the coefficients of integral functions to construct them.
This technique was used to obtain the cut-containing parts of all one-loop
$n$-gluon amplitudes with the helicities arranged in a `split
helicity' configuration.

Very recently, Xiao, Yang and Zhu have presented a different method
for obtaining rational function terms by applying spinor
simplifications together with integrations that target only the
rational terms~\cite{XYZ}.  They have used this to obtain all
the rational terms in the one-loop six-gluon
amplitudes.  Binoth et al.~have also 
presented a diagrammatically-related method~\cite{BinothExtra}.

\subsection{Tree-level Recursion Relations}

On-shell recursion relations have a curious history which did not
foreshadow their widespread applicability.  Motivated by
Witten's conjecture that $\NeqFour$ super-Yang-Mills gauge theory
amplitudes should have a simple structure in twistor space, and by
Brandhuber, Spence, and Travaglini's 
observation~\cite{BST} that this simplicity indeed held
beyond tree-level, for the simplest
maximally-helicity-violating class of one-loop amplitudes, Del Duca
and several of the authors computed the seven-point
next-to-maximally-helicity violating (NMHV) amplitudes
\cite{NeqFourSevenPoint}, one of which was also computed by Britto,
Cachazo and Feng~\cite{CachazoAnomalyBCF7}.  These amplitudes have
three negative helicities, and were expected to lie on a genus-one,
degree-three curve.  Subsequently,
amplitudes with three negative helicities and an arbitrary number of
positive-helicity gluons were 
computed~\cite{NeqFourNMHV,BCFUnitarity}.

The compact forms of seven- and higher-point tree
amplitudes~\cite{NeqFourSevenPoint,NeqFourNMHV} that emerged from
studying one-loop infrared singularity consistency equations, together
with the observations that one-loop $\NeqFour$ super-Yang-Mills
amplitudes are composed solely of box integrals~\cite{BDDK} whose
coefficients may be algebraically determined from products of tree
amplitudes~\cite{BCFUnitarity}, led Roiban, Spradlin and 
Volovich to suggest~\cite{RSVRecursion} the existence
of tree-level on-shell recursion relations.  These recursion relations
were constructed explicitly by Britto, Cachazo and
Feng~\cite{BCFRecurrence}.

Because of the indirect way in which on-shell recursion relations
were obtained, at first it was not clear how widespread their
applicability could be.  However, a simple proof of the tree-level
recursion relations by Britto, Cachazo, Feng and Witten~\cite{BCFW}
followed, based on general factorization properties of tree-level
amplitudes as well as elementary complex variable theory.  The
remarkable generality and simplicity of the proof allowed widespread
application~\cite{MoreTreeRecurResults}, including to
theories with massive particles~\cite{GloverMassive,Massive}
and gravity~\cite{GravityRecursion}.

The proof of the tree-level relations employs a
parameter-dependent shift of two of the external momenta
\begin{eqnarray}
k_j^\mu &\rightarrow& k_j^\mu(z) = k_j^\mu -
       {z\over2}{\sand{j}.{\gamma^\mu}.{l}},\nonumber\\
k_l^\mu &\rightarrow& k_l^\mu(z) = k_l^\mu +
       {z\over2}{\sand{j}.{\gamma^\mu}.{l}} \,,
\label{MomentumShift}
\end{eqnarray}
where $z$ is a complex parameter.  Under this shift, the momenta
remain massless, $k_j^2(z) = k_l^2(z) = 0$, and overall momentum
conservation is maintained.  The $z$ dependence of the momenta
makes the on-shell amplitude, $A(z)$, $z$-dependent as well.

\begin{figure}[tbh]
\vspace{9pt}
\vskip -.5 cm 
\begin{center}
\includegraphics[width=8pc]{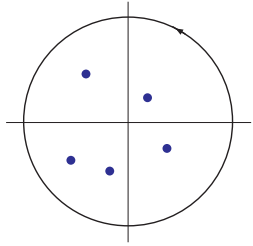}
\end{center}
\vskip -1.3 cm 
\caption{The contour at infinity used for deriving tree-level
recursion relations. The dots represent poles in $A(z)$.}
\label{zplane_treeFigure}
\end{figure}

At tree level, the amplitude, $A(z)$, is a meromorphic rational
function of $z$, so we may exploit Cauchy's theorem to
construct it from its residues. Assuming $A(z)\rightarrow 0$ as
$z\rightarrow\infty$, the contour integral around the circle at
infinity, depicted in \fig{zplane_treeFigure}, must vanish,
\begin{eqnarray}
{1\over 2\pi i} \oint_C {dz\over z}\,A(z)  = 0\,.
\nn
\end{eqnarray}
Using Cauchy's theorem we may evaluate the integral as a sum of 
residues which allows us to solve for the
physical amplitude $A(0)$ in terms of residues on each pole,
\begin{eqnarray}
A(0) = -\sum_{{\rm poles}\ \alpha} \Res_{z=z_\alpha} {A(z)\over z}\,.
\nn
\end{eqnarray}
At tree level, there are many shifts for which $A(z)$ vanishes as
$z\rightarrow\infty$~\cite{BCFRecurrence,BCFW,GloverMassive}.

\begin{figure}[htb]
\centerline{\includegraphics[width=10pc]{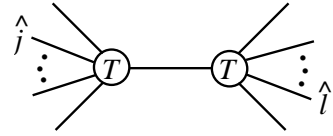}}
\vskip -.7 cm 
\caption{The recursive diagrams at tree level. The `$T$' on each vertex 
signifies an on-shell tree amplitude.}
\label{TreeGenericFigure}
\end{figure}

Each residue comes from factorization in a shifted momentum.
Summing over all residues gives us the tree-level on-shell
recursion relation,
\begin{eqnarray}
A(0) = \sum_{P} \sum_h {A_L^h(z_P)\; A_R^{-h}(z_P) \over K_P^2} \,,
\label{SchematicRecursion}
\end{eqnarray}
where $P$ is the set of ordered partitions of the legs, separating the
two shifted legs $j$ and $l$, and $h$ is the helicity of the
intermediate state.  The lower-point amplitudes $A_L^h(z_P)$ and
$A_R^{-h}(z_P)$ are shifted but with $z$ frozen at the pole,
\begin{eqnarray}
z_P = {K_P^2 \over \sandmm{j}.{\Ksl_P}.{l}} \,.
\nn
\end{eqnarray}
As illustrated in \fig{TreeGenericFigure}, each of the terms in
\eqn{SchematicRecursion} may be given a diagrammatic interpretation,
where the vertices represent lower-point on-shell tree amplitudes.
In them, we
must consider states carrying complex four-momentum, but otherwise
on-shell; transversality conditions as well as overall
four-momentum conservation remain unchanged.

\subsection{Loop-Level Recursion Relations}

\begin{figure}[tbh]
\vspace{9pt}
\vskip -.5 cm 
\begin{center}
\includegraphics[width=8pc]{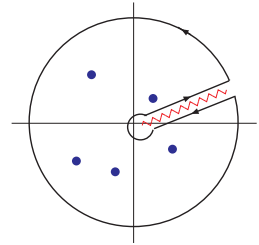}
\end{center}
\vskip -.7 cm 
\caption{A schematic of the contour used for deriving one-loop
recursion relations.}
\label{zplaneFigure}
\end{figure}

\begin{figure*}
\centerline{\includegraphics[width=25pc]{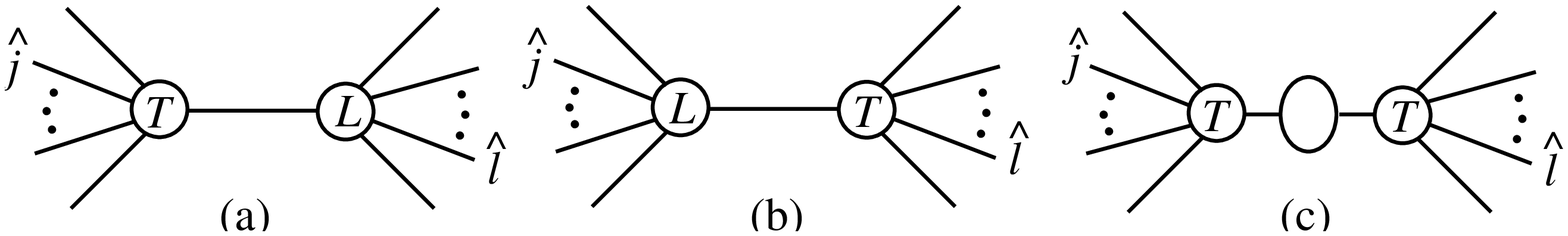}}
\vskip -.7 cm 
\caption{The recursive diagrams at one loop. A `$T$' signifies a tree
amplitude and an `$L$' a loop amplitude. }
\label{LoopGenericFigure}
\end{figure*}

At loop level, we face several issues in constructing
such recursion relations.  The most obvious one is the appearance of
branch cuts, so in addition to the contour used to derive the tree-level
recursion, we will need contours of the form shown in \fig{zplaneFigure}.
We must also deal with spurious singularities, and in some cases,
the non-standard nature of factorization in complex momenta 
(differing from `ordinary' factorization in real momenta). In non-standard
factorization channels (always two-particle ones with like-helicity
gluons), 
double poles and `unreal' poles not
present with real momenta may appear~\cite{OnShellRecurrenceI,Qpap}.  
It is best, and fortunately possible, to avoid these channels in
constructing recursion relations.

This avoidance comes at a price: in general, when choosing shifts 
to avoid non-standard factorizations, the shifted amplitude $A(z)$ may not
vanish as $z \rightarrow \infty$.  The contour integration in
\fig{zplaneFigure} makes it clear
that additional `boundary' contributions arise in this case.
  The approach taken in
ref.~\cite{Genhel}, is to allow for such contributions, and to determine
them using an auxiliary shift and recursion
relation.  Choices for shift momenta with the required properties may be
found in ref.~\cite{Genhel}.

After applying the shift (\ref{MomentumShift}), a loop
amplitude is of the schematic form,
\begin{eqnarray}
A(z)\!\! & = &\!\! \sum \hbox{polylog terms}\nonumber \\ 
  && + \sum_b{ {\rm Res}_i \over z - z_i} + \sum_i a_i z^i \,.
\nn
\end{eqnarray}
Our approach to determining this function
uses the four-dimensional unitarity method to obtain the
polylogarithmic and logarithmic 
terms, on-shell recursion to determine the residues
and an auxiliary recursion relation (when needed) using a different shift to 
obtain the coefficients $a_i$.

In the on-shell bootstrap one first computes the cut-containing
terms.  These will usually contain unphysical spurious
singularities that cancel against the rational functions.  For
reasons of numerical stability it is useful to absorb most of
these into functions that are free of these singularities. We can
construct such functions by adding appropriate rational functions to
the polylogarithmic terms.  For example, we may complete
\begin{eqnarray}
{\ln(r)\over (1-r)^2} \rightarrow 
{\ln(r)\over (1-r)^2} + {1\over 1 - r}\,,
\nonumber
\end{eqnarray}
so that it is singularity-free as $r \rightarrow 1$ ($r$ is
a ratio of kinematic invariants). The use of a `cut completion' also
aids the construction of an on-shell recursion,
since we do not need to compute residues at these unphysical poles.

\begin{figure}[htb]
\centerline{\includegraphics[width=8pc]{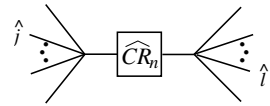}}
\vskip -.7 cm 
\caption{The overlap diagrams corresponding the to different 
physical channels.}
\label{OverlapGenericFigure}
\end{figure}

The recursive contributions, illustrated in \fig{LoopGenericFigure},
are similar to tree-level ones, except that they involve
loop vertices created from loop amplitudes by setting
all cut-containing terms to zero.  
In addition, we have
`overlap contributions' coming from the appearance of physical poles in
the completed-cut terms.  We need to subtract
off this overlap; we simply perform the shift (\ref{MomentumShift}) 
and extract the residues of
the poles in all physical channels.  The correspondence of
these contributions to physical channels again allows us to give them a
diagrammatic interpretation, as illustrated in
\fig{OverlapGenericFigure}. When the
amplitude does not vanish at large values of the shift parameter,
the extra contributions may be computed using an auxiliary recursion
relation as described in
ref.~\cite{Genhel}.

\section{Results}

The bootstrap approach has already been used to obtain the 
rational terms in a variety of new amplitudes:
\begin{enumerate}

\item The finite two-quark $(n\!-\!2)$-gluon amplitudes, with
gluons all of identical helicity~\cite{Qpap}.

\item All one-loop corrections to MHV $n$-gluon
amplitudes~\cite{Bootstrap,FordeKosower,LoopMHV}.

\item All one-loop $n$-gluon amplitudes with three color-adjacent 
negative-helicity gluons and the rest of positive helicity~\cite{Genhel}.
\end{enumerate}
A key feature of our construction of these amplitudes is 
the moderate increase in computational complexity as the number of external
legs increases, in contrast to the explosive growth 
encountered with more traditional methods.

As one example, the rational parts of the six-gluon QCD amplitude
with three color-adjacent negative-helicity gluons may be expressed as,
\begin{eqnarray}
A_{6;1}^{QCD}(1^-, 2^-, 3^-, 4^+, 5^+, 6^+) 
 =  \cg \Bigl[\Cuth_6 + \Remaining_6 \Bigr] \,.
\nonumber 
\end{eqnarray}
where $\cg$ is the constant prefactor that appears in all one-loop
amplitudes, $\Cuth_6$ is the completed cut, which may be obtained
using the four-dimensional unitarity method, and $\Remaining_6$
contains all the remaining rational terms that we are interested in obtaining.
In ref.~\cite{Genhel} these terms were obtained following the
methods outlined above.  They are given by 
a remarkably compact formula,
\begin{eqnarray}
\Remaining_6 = \Remaining_6^a  + \Remaining^a_6 \Bigr|_{\rm flip\; 1} \, ,
\nonumber
\end{eqnarray}
where ${\rm flip\; 1}$ is the flip operation,
\begin{eqnarray}
X(1,2,3,4,5,6)\Bigr|_{\rm flip\; 1} \equiv X(3,2,1,6,5,4) \,, {\rm\ and}
\nonumber
\end{eqnarray}
%
%
\begin{eqnarray}
\Remaining_6^a \hskip -.2 cm &=& \hskip -.1 cm 
 {i\over6} { 1 \over \spb2.3 \spa5.6 \, \spab5.{(3+4)}.2 }
   \Biggl\{ \nn \\
&& \null 
  - { {\spb4.6}^3 \spb2.5 \spa5.6 \over \spb1.2 \spb3.4 \spb6.1 }
  - { {\spa1.3}^3 \spa2.5 \spb2.3 \over \spa3.4 \spa4.5 \spa6.1 }
\nonumber\\
&& \null
 + {{\spab1.{(2+3)}.4}^2 \over \spb3.4 \spa6.1 } \nonumber\\
&& \null \times
      \biggl( { \spab1.{(2-5)}.4\over s_{234} } 
+ { \spa1.3 \over \spa3.4 }
          - { \spb4.6 \over \spb6.1 } \biggl) \nn \\
&& \null
  - { {\spa1.3}^2  ( 3  \spab1.{2}.4 + \spab1.{3}.4 )
     \over \spa3.4 \spa6.1 } \nonumber \\
&& \null
  + { {\spb4.6}^2 ( 3  \spab1.{5}.4 + \spab1.{6}.4 )
     \over \spb3.4 \spb6.1 }
    \Biggr\}
\,. \hskip 1 cm
\nonumber
\end{eqnarray}

\begin{figure}[htb]
\vskip -.6 cm 
\centerline{\includegraphics[width=7pc]{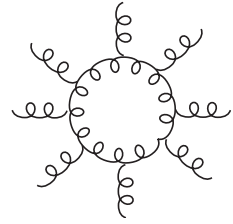}}
\vskip -.8 cm 
\label{eightgluonFigure}
\caption{One of the three million Feynman diagrams for describing
eight-gluon scattering}
\end{figure}

In addition to providing analytic expressions, refs.~\cite{Genhel,LoopMHV}
also record numerical values at sample kinematic points for 
up to eight external
gluons.  A brute-force
computation of eight-gluon amplitudes using standard Feynman diagrams
would require over three million diagrams, 
one of which is displayed
in~\fig{eightgluonFigure}.  Because of the relatively compact nature
of the analytic expressions for the final amplitudes, the numerical
evaluation of these amplitudes is fast.  We do not
anticipate any significant complications arising from round-off error
in numerical evaluation, because of the mild degree
of spurious singularities appearing in the amplitudes.

\section{Outlook}

The developments described above
open several directions that would be interesting
to pursue.
For massive particles inside loops, suitable extensions should be
possible but remain to be developed.  It would be
helpful to have a first-principles derivation of the complex
factorization properties, as well as of the behavior of loop
amplitudes at large values of the shift parameter.  In this regard,
recent papers~\cite{VamanYao} linking tree-level on-shell recursion
with gauge-theory Lagrangians in particular gauges may prove useful.
The unitarity method with $D$-dimensional cuts~\cite{BernMorgan} may
also be useful for developing a formal derivation of properties of
gauge-theory amplitudes.

The on-shell bootstrap approach we have described here has already
established a track record in providing new one-loop amplitudes.
The techniques
we have presented in these talks are systematic and thus should
lend themselves to automation, which will be helpful for dealing with
large numbers of subprocesses.  The method should
carry over to amplitudes with external vector bosons or Higgs
particles, as well as quarks.
It offers a promising approach for attacking the processes needed 
for LHC physics, and we expect that it will see widespread application
towards that goal.

\end{document}